# Constructing a Dataset to Support Agent-Based Modeling of Online Interactions: Users, Topics, and Interaction Networks

Abdul Sittar*    Miha Češnovar†    Alenka Guček*    Marko Grobelnik*

**Abstract**

Agent-based modeling (ABM) provides a powerful framework for exploring how individual behaviors and interactions give rise to collective social dynamics. However, most ABMs rely on handcrafted or parameterized agent rules that are not empirically grounded, thereby limiting their realism and validation against observed data. To address this gap, we constructed a large-scale, empirically grounded dataset from Reddit to support the development and evaluation of agent-based social simulations. The dataset includes 33 technology-focused, 14 climate-focused, and 7 COVID-related aggregated agents, encompassing around one million posts and comments. Using publicly available posts and comments, we define agent categories based on content and interaction patterns, derive inter-agent relationships from temporal commenting behaviors, and build a directed, weighted network that reflects empirically observed user connections. The resulting dataset enables researchers to calibrate and benchmark agent behavior, network structure, and information diffusion processes against real social dynamics. Our quantitative analysis reveals clear topic-dependent differences in how users interact. Climate discussions show dense, highly connected networks with sustained engagement, COVID-related interactions are sparse and mostly one-directional, and technology discussions are organized around a small number of central hubs. Manual qualitative analysis further shows that agent interactions follow realistic patterns of timing, similarity between users, and sentiment change.

Agent-based modeling, online social interactions, information diffusion, network structure analysis, homophily, Reddit discussion networks
=-21pt

## 1 Introduction

Agent-based modeling (ABM) has emerged as a key approach in computational social science for capturing how individual users, each with distinct behavioral rules and network positions, interact to produce collective dynamics. Prior works( [1], [2], [3]) highlighted that ABMs can model heterogeneous agents, complex network structures, emergent phenomena, and support calibration with digital data. More recent studies, such as [4], demonstrate large-scale simulations of information spread across networks, while [5] shows how generative agents based on large language models (LLMs) enhance the realism and adaptability of these models. Despite these advances, a persistent challenge remains: defining realistic agent behaviors based on rich, real-world interaction data and validating simulations against empirical baseline networks.

*E3 Department, Jožef Stefan Institute, Ljubljana, Slovenia. Email: abdul.sittar@ijs.si
†Faculty of Mathematics and Physics, University of Ljubljana, Ljubljana, Slovenia. Email: mc74289@student.uni-lj.si

One of the main challenges in agent-based simulation of online social behavior is the lack of realistic micro-level behavioral baselines. Many ABMs operate at an aggregate level; they rely on parameterized rules for agent posting, commenting, and following, but lack rich empirical interaction data to calibrate these behaviors ( [6]). Calibration of ABMs is hindered by large parameter spaces, randomness in agent behavior, and limited observational data about individual users ( [7]). Further, validation remains difficult: without detailed empirical benchmarks of how users follow, comment, and diffuse content, it is hard to evaluate how well a simulation replicates reality ( [8]). Table 1 summarizes the key differences between traditional agent-based models and recent LLM-driven ABMs with respect to agent design, empirical grounding, validation strategies, and data requirements have been identified in the literature.

Table 1: Comparison of empirical data requirements and grounding approaches in traditional and LLM-driven Agent-based models (ABMs).

| Aspect | Traditional ABMs | LLM-driven ABMs |
| --- | --- | --- |
| **Agent design** | Agents are defined by pre-specified behavioral rules and parameters, often handcrafted or calibrated from small-scale empirical studies [2, 9]. | Agents are instantiated as Large Language Models (LLMs) with adaptive, generative capabilities that emulate human cognition, dialogue, and decision-making [5, 10]. |
| **Data source for calibration** | Relies on surveys, small experimental datasets, or theoretical assumptions; difficult to validate at scale [11]. | Can be grounded in large-scale text corpora (e.g., Reddit, Twitter, news), offering more realistic social context and variability [12, 13]. |
| **Empirical validation** | Validation typically performed via statistical comparison of emergent macro-patterns with real-world data [8, 14]. | Validation can be performed through behavioral alignment with empirical conversational or interactional datasets [15, 16]. |
| **Realism of behavior** | Limited by manually defined rules; difficult to capture nuanced linguistic or emotional behavior [17]. | Enhanced realism through LLM reasoning and dialogue, provided empirical grounding [5, 10]. |
| **Data requirements** | Moderate; structured quantitative inputs [18]. | High; large-scale unstructured text data [12]. |
| **Challenges without empirical data** | Poor external validity [8]. | Risk of hallucinated or unrealistic behaviors [13]. |
| **Empirical grounding benefits** | Improved calibration and credibility [9]. | Reproducible, data-driven synthetic societies [15]. |
| **Example data domains** | Mobility, census, epidemiology [14]. | Social media, conversation logs [10]. |

These limitations become even more pronounced as ABMs increasingly incorporate LLM-driven agents, which simulate human-like dialogue, decision-making, and reasoning. While LLM-based agents can exhibit nuanced linguistic and contextual behavior, their outputs can diverge from real-world social dynamics if not empirically grounded ( [12, 15, 19]). To address this gap, we construct a large-scale, publicly available dataset from Reddit consisting of posts and comments, which provides a behavioral foundation for grounding and validating both traditional and LLM-based agent simulations. We categorize similar users based on content using topic and interaction features, and represent each such collection of users as a single agent. Instead of relying on explicit follow or friendship data (Reddit does not expose) we estimate directed social relations by analyzing temporal commenting patterns between users. Specifically, when users consistently engage with one another's posts over time, we interpret these interactions as indicative of follower–following-like relations. This approach enables us to construct an empirically grounded network of inter-agent connections derived directly from observed social behavior. The resulting dataset can provide a foundation for calibrating, fine-tuning, and benchmarking both rule-based and LLM-based agent behaviors against real social interaction dynamics.

Social media platforms host domain-specific communities in which users exhibit distinct interaction patterns, linguistic styles, emotional expressions, and information diffusion dynamics. Prior studies demonstrate that behavioral and network characteristics vary substantially across topical domains such as politics, health, climate, and technology, influencing how information spreads and how users engage with content [20, 21]. For example, health and crisis related discussions often display heightened emotional intensity and rapid information cascades, whereas technology or science oriented communities tend to exhibit more sustained, expertise driven engagement [22, 23]. Capturing these domain specific differences is critical for agent-based modeling, as agent behaviors calibrated on a single topic may fail to generalize across domains. Multi domain social media datasets therefore enable the construction of heterogeneous agent populations whose behaviors, interaction frequencies, and network structures reflect real variations in online discourse. Such diversity is essential for developing robust simulations that can model domain dependent phenomena such as misinformation spread, polarization, or collective attention shifts, and for validating agent based models against empirical baselines drawn from multiple social contexts [24, 25].

We select Reddit as the social ecosystem for this work for several reasons. Hosting hundreds of millions of users across thousands of thematic communities (subreddits), Reddit provides publicly accessible data for both posts and comments, making it one of the richest open social media platforms for large-scale computational social science. For instance, the Pushshift Reddit dataset contains tens of millions of submissions and comments spanning many years, enabling over-time, cross-community, and network-based analyses ( [26]). Moreover, Reddit covers a wide range of topics (from politics and science to entertainment and niche hobbies) allowing us to categorize users by content and define meaningful agent types. Its public nature and prior usage in content, network, and demographic analyses ( [27,28]) further support its suitability for mapping user categories and deriving follow/comment relationships. While prior work ( [8,29]) highlights technical and coverage limitations in Reddit data, we address these through careful preprocessing and topic-specific filtering. Overall, Reddit offers a scalable, heterogeneous, and openly accessible environment in which we can ground our agent-based simulation approach particularly for developing and validating generative, LLM-powered agents that model realistic social behaviors.

## 1.1 Why empirical datasets are needed for ABMs

The emergence of large language model (LLM)-based agent simulations has underscored a pressing need for empirically grounded datasets that capture real patterns of social interaction, attention, and communication. While recent frameworks (such as EconAgent [30], Generative Agents [10], and AgentSociety [15]) demonstrate that LLM-driven agents can engage in rich, naturalistic dialogues, they largely rely on synthetic or hand-crafted role descriptions and interaction prompts. Without grounding these simulated agents in real-world behavioral distributions, interaction frequencies, and network topologies, such simulations risk producing socially plausible but empirically invalidated dynamics.

In contrast, large-scale social datasets provide the detailed real-world evidence needed to calibrate and validate LLM-based agent societies. Reddit is a particularly valuable source: it contains millions of temporally ordered, user-generated interactions organized around explicit topical communities. These characteristics enable not only the construction of empirically derived social graphs but also the extraction of behavioral baselines such as posting frequency, conversational persistence, and topical engagement intensity that can directly inform agent-level decision parameters in ABMs.

By providing a dataset where users are categorized into topic-aligned agent types and linked via empirically inferred following relationships, we enable a new class of data-aligned simulations. Such datasets allow modelers to (1) fine-tune LLM agents toward realistic social behaviors observed in actual online ecosystems, (2) benchmark the emergent dynamics of simulated societies against observed network metrics, and (3) explore counterfactual or policy experiments in environments constrained by real human behavioral patterns rather than synthetic assumptions. This empirical grounding thus forms the essential bridge between generative simulation and measurable social reality, allowing agent-based models to move from narrative realism toward scientific reproducibility.

## 2 Contributions

Our key contributions are as follows:

- We propose an empirically grounded pipeline for building agent-based simulation inputs from large-scale social media data. The pipeline categorizes agents based on their observed behavior. It infers interactions from user activity patterns. It then constructs interaction networks suitable for simulation.

- We release a large-scale dataset derived from Reddit across three topical domains: technology, climate, and COVID-19. The dataset maps users to content-based agent types. It also includes a directed, weighted interaction network inferred from temporal commenting behavior.

- We provide benchmark analyses of user activity and network structure. We also study triadic closure and how it evolves over time. These analyses can be used to calibrate and validate agent-based simulations.

## 3 Related Work

### 3.1 Agent-based modeling and social simulation

Agent-based modeling (ABM) has become a key approach in computational social science, allowing researchers to study how individual behaviors lead to large-scale social patterns ( [2], [31]). Early ABMs, such as the Schelling segregation model and Axelrod's cultural dissemination model, showed how simple local rules can produce complex outcomes. In online settings, ABMs have been used to study the spread of emotions, influence, and information ( [32], [33], [34], [4]). While traditional ABMs rely on hand-crafted rules, recent approaches use large language models (LLMs) or data-driven techniques to create more realistic agents that can generate behavior informed by real data ( [5], [15]). These advances help bridge the gap between synthetic simulations and real-world social dynamics. Traditional ABMs face persistent limitations, including simplistic decision logic, lack of empirical calibration, and limited scalability or validation ( [17], [35]). Manually designed agents often ignore evolving goals, cognitive context, and heterogeneity factors crucial for modeling realistic online interactions. Recent work has sought to address these issues through data-driven and learning-based agents, including LLM-powered agents ( [5]), moving toward empirically validated simulations. Our work contributes to this direction by constructing a Reddit-based dataset where users are categorized into agent types derived from content, and inter-agent relationships are inferred from comments and follows. This provides a data-aligned foundation for simulating and evaluating social dynamics, bridging the gap between manually designed and data-grounded agent-based models.

## 3.2 Data-driven simulation and fine-tuning of agents

Recent research has leveraged large language model based agents and empirical data to increase the realism of agent-based simulations. [12] survey the space of LLM-driven social simulation and categorize simulation types from individual to societal scale. The work by [5] introduces LLM-empowered agent frameworks for social network simulation, while [15] build a large-scale system with 10k+ agents interacting in a rich societal environment. [13] emphasis the necessity of grounding agent profiles in real-world data rather than synthetic ones, and [16] use LLM agents in simulated social media to test influence mechanisms. These works collectively point to the potential of combining real interaction data with agent simulation. However, there remains a gap: existing simulations often lack fine-grained mapping of user content interaction data (posts, comments, follows) into agent categories and calibrated inter agent relations. Our work addresses this by collecting large-scale user content/comment data from Reddit, converting users into content-based agent categories, estimating inter-agent relations from comment behavior, and releasing a dataset to enable fine-tuning of agents and evaluation of simulation realism.

## 3.3 Reddit and online network analysis

Recent research has explored Reddit as a platform for network analysis, community structure, and information diffusion. ( [36]) studied the tree structures of Reddit comment threads to understand how global and local features shape discussion depth, width, and size. [37] investigated community identity and its effect on user engagement and retention across multiple subreddits. [38] examined inter-community interactions, mobilizations, and conflicts, highlighting the complex dynamics of Reddit's multi-community environment. Other works, such as ( [39]) and ( [40]), have constructed subreddit-level networks to study cross-community relations and information flows. Additionally, studies have examined temporal dynamics within user sessions, showing phenomena like performance deterioration over active sessions ( [41]). These prior studies provide a strong foundation for network, content, and temporal analyses, yet few integrate content-based user categorization, follower/following relationships, comment interactions, and agent-based simulation, which is the focus of our work. Beyond Reddit, systems such as MIDDAG focus on social media–driven information diffusion, modeling how COVID-19 news propagates through user engagement, event tracking, and diffusion forecasting [42–45].

## 3.4 Bridging real and synthetic social data

A key challenge in social simulation is ensuring that synthetic or agent-based models meaningfully reflect real-world network and behavioral patterns. Several prior works have addressed this: for example, [46] compared synthetic contact networks with empirical networks, demonstrating the divergence in structural metrics. Similarly, [47] derived agent behavior from an online community and validated an ABM by reproducing the observed network. On the network-generation side, heuristic methods such as personality or compatibility based synthetic networks have attempted to close this realism gap [18]. More broadly, methodological reviews emphasize that combining Social Network Analysis (SNA) and Agent-based modeling (ABM) offers a powerful route to bridge empirical data and simulation frameworks [48]. On the cutting edge, the recent work by [5] integrates generative language-model agents with network simulation and validates against real social media behavior. Our work builds on this trajectory by collecting a large-scale empirical dataset of users, posts, comments from Reddit, converting users into agent categories, estimating inter-agent relations via comment behavior, and releasing a dataset and simulation-ready framework. Unlike many prior studies (which either simulate networks without detailed content-based agent categorization or adapt small/closed communities), our approach combines content, network structure, temporal dynamics, and large-scale publicly accessible data. This dataset can be used to fine-tune agents and benchmark synthetic vs. real discussion dynamics.

# 4 Methodology

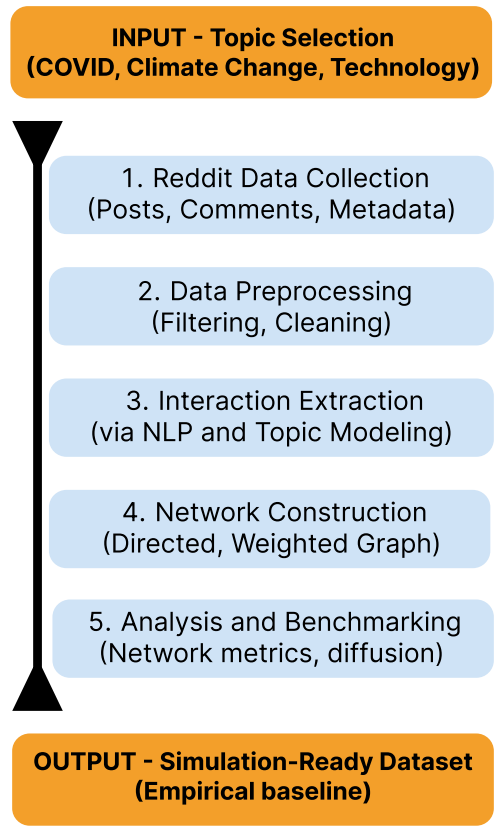

Figure 1: Methodological pipeline for constructing an empirically grounded agent-based simulation dataset from Reddit interactions.

Our goal is to construct an empirically grounded foundation for agent-based social simulations using real online interaction data from Reddit. Figure 1 presents the end-to-end methodological pipeline, consisting of five sequential stages: data collection, preprocessing, user categorization and interaction extraction, network construction, and analysis and benchmarking.

We begin by collecting publicly available Reddit data through the Pushshift API or PRAW (Python Reddit API Wrapper). It contains many subreddits, we filtered subreddits named as COVID-19 (2020-2024), technology (2020-2024), and climate changes (2009-2024 ). The dataset includes posts, comments, timestamps, subreddit identifiers, and anonymized user IDs (see Table 2). This data captures both content (textual data) and structure (who interacts with whom), providing a rich foundation for modeling social behavior (Black see the Github Link [1]). Black The dataset is publicly available on Zenodo: https://zenodo.org/records/18082502 [2] )

Raw Reddit data require extensive preprocessing to ensure reliability and reproducibility. Table 2 summarizes the effects of each preprocessing step across the technology, climate, and COVID-19 domains. We:

- Remove deleted or bot accounts and filter out low-quality or spam content.
- Standardize timestamps and metadata.
- Optionally select specific subreddits or topics for focused studies (e.g., politics, gaming, or science).

---

[1] https://github.com/abdulsittar/Social-Graph-Inference-Reddit.git
[2] https://zenodo.org/records/18082502

Figure 2: A visualization of all agents: 33 technology-focused, 14 climate-focused, and 7 COVID-related agents. The size of each bubble reflects the agent's level of activity, measured as a weighted sum of posts and comments.

- Perform language detection and retain only English-language posts, unless multilingual analysis is intended.

- These steps ensure a clean, consistent dataset suitable for further semantic and structural analysis.

| ID | Domain | Total Posts | Total Comments | Preprocessing Step |
|---|---|---|---|---|
| 0 | Technology | 2,207,316 | 30,804,312 | Raw data (no filtering) |
| 1 | Technology | 284,294 | 628,841 | Bot removal, topic noise filtering, truncated to first 10 comments |
| 2 | Technology | 135,156 | 152,516 | User activity thresholding (min. interactions) |
| 3 | Technology | 111,243 | 151,060 | Removal of deleted/removed posts and comments |
| 4 | Technology | 111,243 | 151,060 | Stylistic feature extraction (keywords) |
| 5 | Technology | 111,243 | 151,060 | Stylistic and emotional feature enrichment |
| 6 | Technology | 111,243 | 151,060 | Inferred follower–following relationships |
| 0 | Climate | 226,766 | 1,914,565 | Raw data (no filtering) |
| 1 | Climate | 116,710 | 341,031 | Bot removal, topic noise filtering, truncated to first 10 comments |
| 2 | Climate | 101,662 | 121,139 | User activity thresholding (min. interactions) |
| 3 | Climate | 95,997 | 121,139 | Removal of deleted/removed posts and comments |
| 4 | Climate | 95,997 | 121,139 | Stylistic and emotional feature enrichment |
| 5 | Climate | 95,997 | 121,139 | Agent profile construction (keywords and emotion) |
| 6 | Climate | 95,997 | 121,139 | Inferred follower–following relationships |
| 0 | COVID | 262,912 | 3,300,050 | Raw data (no filtering) |
| 1 | COVID | 98,067 | 445,194 | Bot removal, topic noise filtering, truncated to first 10 comments |
| 2 | COVID | 83,767 | 179,951 | User activity thresholding (min. interactions) |
| 3 | COVID | 83,767 | 156,817 | Removal of deleted/removed posts and comments |
| 4 | COVID | 83,767 | 156,817 | Stylistic and emotional feature enrichment |
| 5 | COVID | 83,767 | 156,817 | Agent profile construction (keywords and emotion) |
| 6 | COVID | 83,767 | 156,817 | Inferred follower–following relationships |

Table 2: Dataset preprocessing pipeline across technology, climate, and COVID domains.

To represent heterogeneous agent populations, we categorize users into agent types based on content and behavior. Topic modeling techniques such as BERTopic, and sentence-transformer clustering are applied to users' aggregated posts and comments. Each cluster is then manually assigned an agent category reflecting their primary thematic or behavioral orientation (e.g., informative, opinionated, and supportive). This step grounds agent heterogeneity in empirical data rather than synthetic assumptions.

Next, we infer directed social interactions from comment or reply structures. If user A comments on user B's post or reply, we record a directed edge A → B. Each edge is weighted by interaction frequency. Temporal ordering is preserved to allow later dynamic modeling. The resulting edges encode the social topology of user interactions observed in the platform.

We aggregate the extracted edges into a directed, weighted graph where nodes represent agents and edges represent empirical interactions. The graph captures realistic degree distributions, clustering, and community structures observed in online environments. Structural features such as reciprocity, triadic closure, and modularity can be analyzed to benchmark the fidelity of the empirical social fabric.

Finally, we compute network and behavioral metrics (e.g., degree centrality, assortativity, clustering coefficient) and simulate diffusion dynamics (e.g., information spread, opinion propagation) to compare empirical patterns with those generated by simulated agents. These analyses serve two purposes: 1) They validate that the constructed dataset reflects authentic social dynamics. 2) They provide empirical baselines for calibrating and evaluating future agent-based models.

The outcome of this pipeline is a simulation-ready dataset consisting of: 1) An agent–agent interaction network grounded in real data., 2) Agent category distributions representing behavioral diversity (see Fig. 2)., 3) Benchmark metrics for comparing synthetic simulations to observed online phenomena. This methodology bridges the gap between purely synthetic agent-based modeling and empirically validated social simulation, enabling more realistic exploration of online social systems.

## 4.1 Estimating following relationships from user interactions

To infer implicit following relationships between Reddit users across topic-specific discussions (e.g., technology, climate, or COVID-19), we rely on the temporal consistency of user-to-user interactions. Specifically, we treat a user's comments directed toward another user as potential indicators of attention or social linkage. We formalize this interaction-based following inference as shown in Eq. 1.

Let $C_{u \to v}(t)$ denote the number of comments user $u$ makes in reply to user $v$ during a given time window $t$. We define the following relationship score between $u$ and $v$ over $n$ consecutive observation periods as

$$F_{u,v} = \begin{cases} 0, & \text{if } \sum_{t=1}^{n} \mathbb{I}[C_{u \to v}(t) > 0] < 2, \\ \text{"maybe"}, & \text{if } \sum_{t=1}^{n} \mathbb{I}[C_{u \to v}(t) > 0] = 2, \\ \text{"for sure"}, & \text{if } \sum_{t=1}^{n} \mathbb{I}[C_{u \to v}(t) > 0] \geq 3, \end{cases} \quad (1)$$

where $\mathbb{I}[\cdot]$ is the indicator function. In words, when a user comments on another's posts or replies consistently over time (at least three times across separate periods) we consider this a confirmed (*for sure*) following. One or two consistent interactions are treated as a potential (*maybe*) following.

This rule-based inference provides a simple yet interpretable approximation of directed ties in social conversation networks where explicit "follow" actions are not observable.

This approach aligns conceptually with prior studies that model social link formation as a function of behavioral and structural signals. For instance, a longitudinal study of Twitter follow predictors [49] demonstrated that message content, social behavior, and network structure jointly explain follow formation over time, suggesting that temporal consistency captured here through repeated directed commenting serves as a key predictor of emerging ties. Similarly, work on predicting rising follower counts using profile information [50] emphasized discoverability features and user attributes, while our formulation abstracts such structural and semantic factors into an observable behavioral pattern: interaction persistence.

Several studies on reciprocity and link evolution further motivate our temporal framing. Research on predicting reciprocity and triadic closure [51, 52] showed that one-way (parasocial) links often evolve into mutual (reciprocal) ties as users exchange repeated interactions. Our model parallels this by distinguishing between "maybe" and "for sure" followings—interpretable as a behavioral progression from parasocial attention to stable reciprocal engagement. Studies such as [53] on parasocial versus reciprocal relationships highlight that user attributes, network degrees, and behavioral similarity shape whether a directed connection becomes mutual; our rule operationalizes these dynamics in the absence of explicit follow data.

More recent research expands follow and reciprocity prediction to dynamic and generative models. For example, dynamic stochastic block models integrating reciprocity [54] capture how communities evolve and reciprocate connections over time. Our approach offers a simple, data-driven alternative that captures similar temporal patterns using discrete interaction windows. Likewise, link-prediction frameworks that combine heuristic and representation-learning features [54] achieve high predictive accuracy but often sacrifice interpretability. In contrast, our method maintains interpretability while revealing consistent interaction-based follow patterns within evolving discussion networks.

Finally, studies on popularity and follow-back behaviors [55, 56] show that reciprocal following and early adopter effects drive visibility and community growth. Our inference of "for sure" ties can be viewed as identifying those stable, mutually attentive connections that correspond to socially reinforced or influence-generating relationships in other platforms.

Table 3: Manual evaluation questions used to assess agent interactions and emergent social behavior

| Q# | Evaluation Question | Aspects Considered | Analytical Perspective |
|---|---|---|---|
| 1 | Do agents generate meaningful commentary on posts over time? | Post and comment timing | Cascade analysis; information flow and diffusion |
| 2 | Are interacting agents similar in terms of their profiles? | Agent profile characteristics | Engagement life cycle |
| 3 | Why do two agents interact more frequently after a specific interaction? | Interaction triggers and context | Content interaction patterns |
| 4 | Does the discussion extend among agents with similar characteristics? | Similarity between interacting agents | Homophily |
| 5 | Do sentiment patterns align with known behavioral dynamics? | Emotion evolution over time | Sentiment analysis |
| 6 | Do interactions exhibit multi-turn conversational exchanges? | Presence of back-and-forth replies | Conversational dynamics |
| 7 | Does the topic of discussion remain same across all chains? | Similar topic | Topic modeling and topic drift |

While prior research largely focuses on predicting follow or reciprocal edges on platforms such as Twitter or Google+, our model extends these ideas to environments (e.g., Reddit). By formalizing temporal consistency in directed commenting as a proxy for latent following, we bridge behavioral studies of interaction persistence with network-theoretic research on reciprocity and link formation. This provides a unified, interpretable basis for quantifying user attention and community evolution across topic-specific social ecosystems.

# 5 How realistic are the aggregated agents and the estimation of network connectivity?

## 5.1 Qualitative Evaluation

The qualitative evaluation focuses on interaction timing, cascade structure, homophily, and sentiment evolution, drawing on empirically grounded interaction chains (see Fig. 3 and Table 3). To complement the quantitative structural, network, and behavioral analyses, we conducted a manual qualitative evaluation of selected agent interactions. The objective of this evaluation is to assess whether the generated agent behavior exhibits plausible human-like conversational dynamics, semantic coherence, and social patterns that are difficult to fully capture using automated metrics alone. The evaluation focuses on interaction timing, cascade structure, profile similarity, interaction escalation, homophily, and sentiment evolution, drawing directly from empirically grounded interaction threads.

In total, we manually evaluated 105 interaction chains across three domains: Technology, Climate Change, and COVID-19. To construct these chains, we first performed topic modeling on each generated post and its associated comments. Pairwise comparisons were then conducted between posts and comments using cosine similarity over their topic representations. When the cosine similarity between two items exceeded a threshold of 0.1, we defined this relationship as a semantic connection. Connections were directed from the earlier node to the later node based on posting timestamps, yielding temporally ordered interaction graphs. If a node exhibited multiple

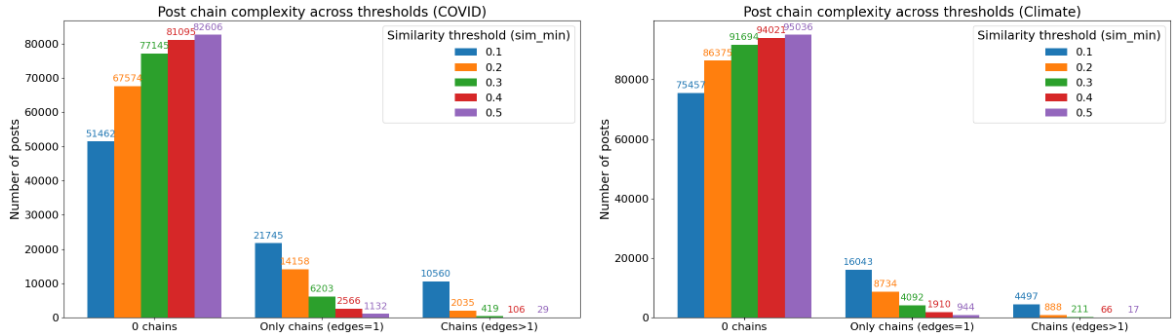

(a) COVID domain: distribution of interaction chains used for qualitative evaluation.

(b) Climate domain: distribution of interaction chains used for qualitative evaluation.

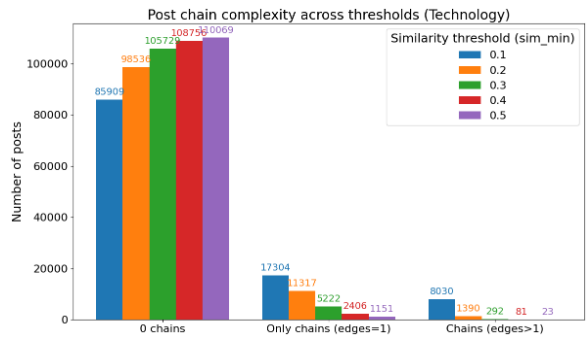

(c) Technology domain: distribution of interaction chains used for qualitative evaluation.

Figure 3: Distribution of post-level interaction chain complexity across similarity thresholds (0.1–0.5). For each threshold, posts are grouped by whether they form no interaction chains, only single-edge chains (length = 1), or also longer chains (length > 1). Counts indicate the number of posts in each category.

outgoing semantic connections, the corresponding interaction chain was duplicated such that each resulting chain contained only a single incoming and a single outgoing connection per node. This normalization step ensured that each chain represented a linear conversational trajectory rather than a branching structure. The resulting chains were ranked by length, and for manual qualitative evaluation we selected the 35 longest chains per domain, resulting in a total of 105 evaluated chains.

**1. Information flow and cascade dynamics:** Manual analysis of post–comment timelines reveals that agent interactions follow realistic temporal patterns. Initial posts typically trigger a few of replies shortly after publication, followed by declining engagement over time.

**2. Engagement life cycle:** Manual comparison of agent profiles including dominant keywords, thematic focus, stylistic and emotional features shows that agents who interact repeatedly tend to share overlapping topical interests and communicative styles. Agents with similar profiles engage more frequently and sustain interactions over longer periods.

**3. Content interaction patterns:** Across multiple evaluated chains, a clear escalation is observed following the first meaningful exchange between agents. Once two agents engage directly (e.g., via an explicit reply), subsequent interactions become more frequent and more targeted.

**4. Homophily and social similarity:** Manual analysis confirms that agents tend to interact more with others who are similar to them. Similarity appears in topic interest, expression opinions, and communication styles. This pattern is visible in repeated one on one interactions.

**5. Sentiment dynamics:** Manual sentiment inspection reveals emotionally coherent trajectories across interaction chains. Neutral or informational exchanges tend to remain emotionally stable, while argumentative or controversial discussions often show increasing emotional intensity, including frustration or negativity.

**6: Conversational dynamics:** Extended multi-turn conversations are rare. Most chains consist of only a few turns (typically up to 6-7 comments), often dominated by one or two agents. This limited depth may partly result from agent grouping strategies and similarity-based interaction rules.

**7. Topic modeling and topic drift:** Topics within interaction chains remain relatively stable. There is little evidence of significant topic drift within short-lived chains.

**Evaluation summary:** The analysis shows that interaction chains are generally short and limited in scope. Most chains consist of up to 6–7 comments and typically unfold over a few days. Two dominant interaction patterns emerge: one where a user states an opinion and others argue or respond critically, and another where a user asks for similar experiences and others reply in a question–answer or discussion format. About half of the chains involve only a single agent. When multiple agents participate, they are usually very similar in emotional tone and writing style. Often, only one or two aggregated agents dominate the interaction throughout the entire lifecycle of a chain. Multi-turn or extended discussions are rare. When they do occur, they tend to stay focused on the same topic and are usually driven by the same two agents. This limited conversational depth may be influenced by how agents were grouped by topic and similarity. Agent interactions strongly reflect homophily: agents mostly interact with others who are similar to themselves in tone, emotion, and style. Emotional dynamics within chains often show a progression, starting from a neutral tone and gradually intensifying toward stronger emotions such as anger, fear, or joy. These emotional trajectories generally align with both the topic of discussion and the typical emotional profiles of the agents involved, and metadata-based sentiment analysis mostly confirms these patterns. Finally, network growth effects are difficult to identify clearly. However, there is some indication that agents who comment on another agent's post often end up following that agent, suggesting a weak but observable link between interaction and network connections.

Overall, the manual qualitative evaluation shows that agent interactions reflect key characteristics of realistic online behavior. These include bursty and cascading interaction patterns, sustained engagement driven by profile similarity, increased interaction after initial contact, homophilic alignment, and believable sentiment changes across multi-turn discussions. Importantly, these observations align well with the quantitative metrics used in our analysis and help explain them. Together, these results support the conclusion that the generated agent dataset captures not only realistic network structure but also meaningful semantic, emotional, and social behaviors. As a result, the manual evaluation adds an important layer of interpretability and ecological validity, increasing confidence in the dataset's suitability for agent-based modeling and social simulation studies.

## 5.2 Quantitative Evaluation

### 5.2.1 In-degree centrality: identifying influential nodes

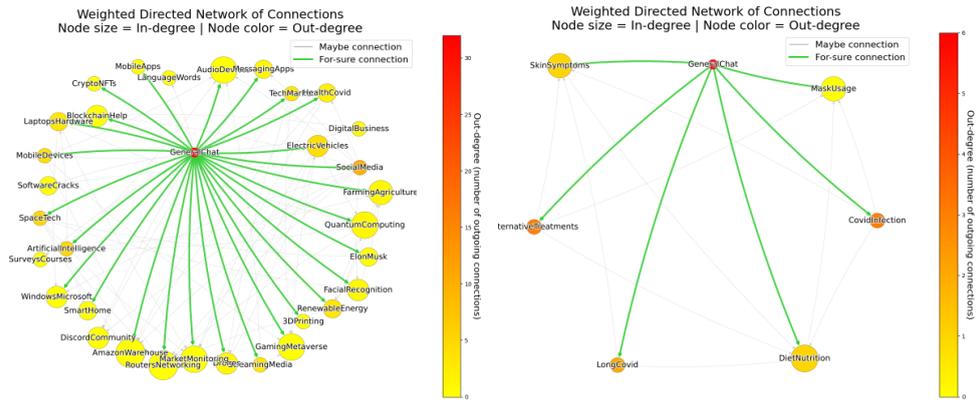

(a) Technology domain: top nodes by in-degree at 0.01% coverage threshold (maybe relationships).

(b) COVID domain: top nodes by in-degree at 0.01% coverage threshold (maybe relationships).

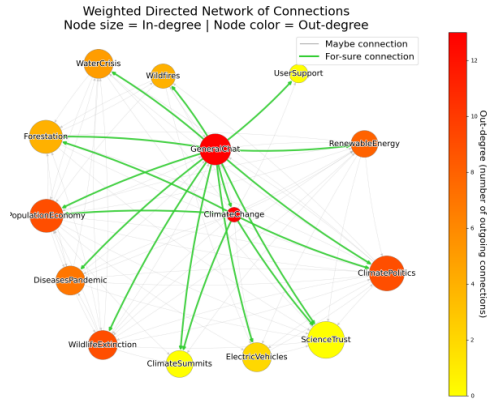

(c) Climate domain: top nodes by in-degree at 0.01% coverage threshold (maybe relationships).

Figure 4: Visualization of inferred following relations at a 0.01% similarity threshold across the Technology, COVID, and Climate domains. Edges indicate "for sure" (green) and "maybe" (grey) connections between posts.

Figure 4a highlights the most prominent users or topics in terms of in-degree, indicating those followed by a large number of others at the 0.01% coverage threshold within the "maybe" category (less certain but still notable connections). The edges represent potential follow relationships estimated from user behaviors such as commenting or engagement patterns. In-degree here measures how many unique users appear to follow or interact with each node, reflecting its popularity or centrality within the network. At the top of the ranking is *DiscordCommunity*, which stands out significantly with the highest in-degree count, suggesting it is the most followed or most connected topic at this detection level. It likely represents a highly active or engaging domain that attracts attention across various user groups.

The network graph (Fig. 4b) displays the top users or topics most followed by others in the COVID domain, based on in-degree at the 0.01% coverage threshold under the "maybe" relationship category. In this context, in-degree reflects the number of inferred connections directed toward each topic, capturing which subjects attract the most user engagement or attention. The most prominent node is *MaskUsage*, which shows the highest in-degree count, suggesting that discussions or user interactions surrounding mask-related behavior were central and widely referenced. Close behind are *DietNutrition* and *SkinSymptoms*, both exhibiting strong connectivity—indicating that users frequently interacted with content linked to health maintenance, nutrition, and symptom tracking during the pandemic period. Further down the ranking are *LongCovid*, *AlternativeTreatments*, and *CovidInfection*, which—while less dominant—still display significant interaction counts. Their presence highlights how extended health concerns and debates about treatment efficacy contributed to sustained user engagement. Collectively, these results show that health-related subtopics, rather than policy or logistics, dominated attention in the COVID discourse at this inferred level. The use of the "maybe" network captures a broader behavioral footprint, identifying emerging areas of collective focus where connections may be exploratory or indirect, but nonetheless indicative of influence within the overall information ecosystem.

The network graph (Fig. 4c) presents the top users or themes with the highest inferred in-degree values within the Climate domain at the same 0.01% coverage threshold ("maybe" category). Here, in-degree measures how many other nodes appear to reference or engage with each topic, revealing which areas serve as focal points in climate-related discourse. The leading node, *ScienceTrust*, stands out markedly as the most followed or referenced topic, underscoring the centrality of scientific credibility and public trust within environmental conversations. This is followed by *DiseasesPandemic* and *GeneralChat*, which likely reflect overlap between climate discussions and broader societal or health concerns, as well as general spaces of user interaction. Mid-tier nodes such as *Forestation*, *WildlifeExtinction*, and *ClimatePolitics* indicate active engagement with ecological and political aspects of climate change, suggesting a mix of activism, awareness, and debate-driven interactions. Meanwhile, *Wildfires*, *ElectricVehicles*, *WaterCrisis*, and *UserSupport* occupy the lower range of the top ten but still maintain substantial in-degree values, reflecting persistent but more specialized interest clusters. Overall, the Climate network exhibits a diverse structure of attention, with topics spanning science, policy, and sustainability. The inferred ("maybe") connections reveal a broad, interconnected community of discussion that, while diffuse, collectively emphasizes trust in science and multi-sectoral environmental awareness.

Across all domains, in-degree centrality reveals distinctive influence structures: the Tech network centers on functional and utility-driven hubs, COVID discussions highlight behavioral and health-oriented engagement, and climate discourse underscores the role of scientific trust and interdisciplinary connectivity.

### 5.2.2 Out-degree centrality: identifying active or exploratory nodes

The users or topics with the highest number of outgoing connections are presented in Figure 4a, reflecting strong outward engagement at the 0.01% coverage level for the "maybe" category. In a network sense, out-degree represents how many outgoing links or inferred follows each user or topic initiates—identifying the most active or exploratory nodes in forming connections with others. This visualization captures estimated relationships derived from user interaction data (such as commenting or co-discussion patterns), focusing on potential rather than fully con-firmed connections. The chart is dominated overwhelmingly by *GeneralChat*, which exhibits a far higher out-degree count than any other node, suggesting it is the most broadly connected or interactive topic in the dataset. Content or users associated with *GeneralChat* appear to engage with a wide variety of other topics, effectively acting as a central hub or bridge within the network. After *GeneralChat*, the next most connected topics—*SocialMedia*, *SpaceTech*, *ElectricVehicles*, *MobileDevices*, and *TechMarket*—have much smaller counts, indicating significantly fewer outgoing links. These nodes still play active roles in forming connections but at a much lower scale compared to *GeneralChat*.

In Figure 4b, the distribution of out-degree identifies the users or topics that most frequently initiate interactions within the COVID-related discussion network under the 0.01% "maybe" threshold. Out-degree in this context represents the number of other users or topics that a node follows, mentions, or interacts with—capturing the extent of its outward engagement. The chart reveals that *GeneralChat* dominates the network, exhibiting the highest out-degree by a significant margin. This suggests that it functions as a highly interactive or bridging node, engaging across numerous COVID-related topics. Following it is *CovidInfection*, which also maintains a strong presence, implying active interaction with various subtopics in the COVID discourse. *AlternativeTreatments* and *LongCovid* occupy mid-tier positions, indicating moderate but targeted engagement—these nodes likely focus on more specific or niche discussions within the larger COVID context.

Figure 4c depicts the most active initiators of connections in the Climate-related network, as measured by out-degree under the 0.01% "maybe" threshold. Here, out-degree reflects how broadly each node reaches out to others, signifying active participation and cross-topic engagement within environmental discussions. In this visualization, *ClimateChange* overwhelmingly dominates, with an out-degree count far surpassing all other nodes. This indicates that *ClimateChange* acts as a major connector and initiator of discussions across the climate discourse, serving as the primary hub linking multiple thematic areas. The next most connected node, *GeneralChat*, also demonstrates significant outward engagement, though still at a much lower scale—suggesting it facilitates broader, cross-domain communication. Below these top two, nodes such as *RenewableEnergy*, *PopulationEconomy*, *WildlifeExtinction*, and *ClimatePolitics* exhibit moderate out-degree values, representing more specialized yet active participants within their respective niches. Toward the bottom of the chart, *DiseasesPandemic*, *WaterCrisis*, *ElectricVehicles*, and *Forestation* maintain low out-degree counts, implying either topic-specific isolation or limited engagement breadth.

Across domains, out-degree centrality highlights different modes of activity: the Tech network is driven by exploratory general-interest nodes, COVID discourse shows limited but health-focused outward engagement, and Climate discussions reveal centralized thematic outreach dominated by broad, unifying concepts like climate change.

### 5.2.3 Triadic closure dynamics: network cohesion over time

Next we are interested in the evaluation of triadic closure dynamics across three domains—Technology, COVID, and Climate—at a 0.01% coverage threshold. Figure 5 compares the evolution of triadic closures over time, with each panel showing the number of closed triads aggregated in six-month intervals. Triadic closure is a key indicator of network cohesion, representing how fragmented or tightly interconnected a community becomes as indirect relationships (A–B–C) turn into direct ones (A–B–C–A).

Across all three domains, the blue solid line ("All edges = for sure + maybe") consistently shows an upward trend, indicating an overall increase in network interconnectedness. However, the rate and magnitude of this growth vary by domain.

In the Tech domain (Fig. 5a), triadic closure rises sharply after 2020, growing from near zero to roughly 80 by 2025. This rapid increase suggests an intensifying overlap among users or topics—possibly reflecting how technological discussions and influencers become progressively

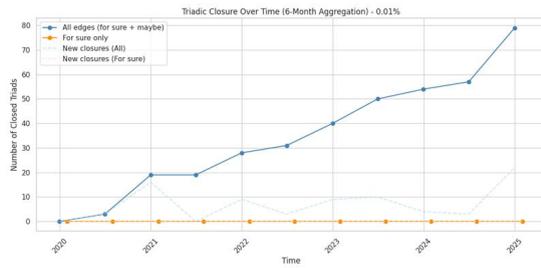
(a) Technology domain: triadic closures over time.

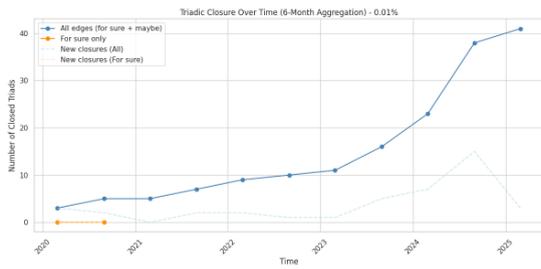
(b) COVID domain: triadic closures over time.

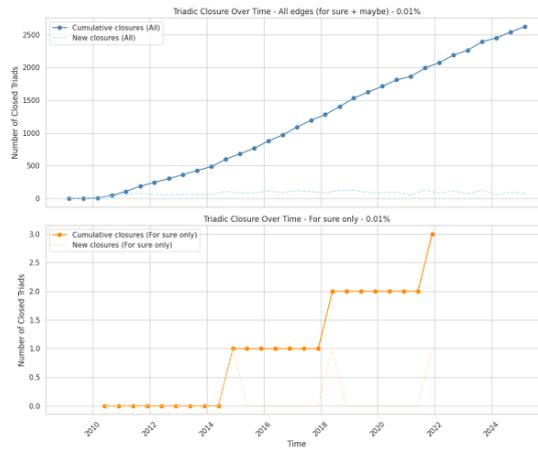
(c) Climate domain: triadic closures over time.

Figure 5: Comparison of triadic closures over time across all 3 domains for all connections (blue) and only "for sure" connections (orange).

interlinked over time. The corresponding dashed blue line ("New closures") fluctuates but remains generally positive, implying steady formation of new triangles. In contrast, the orange ("For sure only") lines remain flat near zero, showing that confirmed, high-confidence ties contribute minimally to this growth; most of the observed cohesion stems from tentative or inferred relationships.

A comparable but more gradual trend is observed in the COVID domain (Fig. 5b), where the number of closed triads increases slowly, reaching approximately 40 by 2025. This indicates moderate network consolidation—users and topics become more interconnected, but the process is less intense than in the Tech domain. As with Tech, "for sure" closures are nearly absent, implying that while inferred connections are growing, strong mutual links between key nodes remain limited. This could reflect the short-term or event-driven nature of COVID-related discourse, where users cluster temporarily around evolving issues.

In contrast, the Climate domain (Fig. 5c) displays a more gradual yet sustained increase in triadic closure, particularly in the cumulative plots. The total number of closed triads continues to rise steadily across years, indicating long-term structural growth in network cohesion. Interestingly, small but visible steps appear in the "for sure only" lines, suggesting that although fewer in number, high-confidence connections do occasionally form and persist. This pattern may point to a more stable, enduring community structure typical of ongoing policy or scientific discussions, where relationships are built over longer periods rather than through rapid bursts of interaction.

Overall, comparing across domains, Tech exhibits the fastest and largest growth in inferred connectivity, COVID shows moderate but short-lived clustering, and Climate reflects slower yet more stable network formation. The consistent absence of strong triadic closures across all domains underscores that much of the apparent interconnection arises from weaker or uncertain ties, hinting at broad but not deeply cohesive community structures.

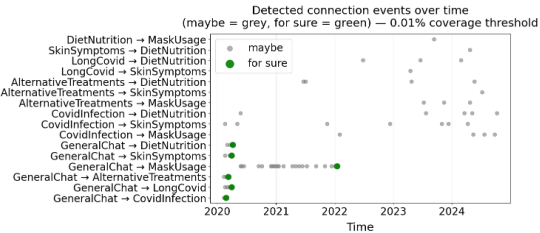

Figure 6: Technology domain: detected following events over time at 0.01% threshold. Detected following events over time. Grey dots indicate *maybe* followings (two or fewer consecutive user interactions), while green dots denote *for sure* followings (three or more consistent directed interactions) inferred from commenting behavior. The full figures for all three domains have been provided in the appendix.

### 5.2.4 Cross-domain comparison of inferred following dynamics

Figure 6 compares inferred following events across three topical domains—technology, COVID-related health discussions, and climate/environmental topics—using the same consistency-based detection rule defined in Equation 1. In all cases, directed commenting patterns are treated as implicit signals of "following" relationships, where the temporal persistence of such interactions determines their strength.

As shown in Fig. 6a, most strong (*for sure*) connections emerge during the 2020–2021 period. This concentration suggests a surge of interaction intensity during those years, particularly around discussions involving `GeneralChat`, `SocialMedia`, `HealthCovid`, and technical subtopics such as `QuantumComputing` and `ArtificialIntelligence`.

Table 4: Comparative interpretation of interaction networks by topic.

| Topic | Key Characteristics |
|---|---|
| Climate | Dense, highly clustered, moderately reciprocal discussions with sustained interaction and shared attention. |
| COVID-19 | Sparse, low-reciprocity interactions dominated by one-way information exchange. |
| Technology | Fragmented, hub-driven structure with strong topical segmentation and minimal conversational reciprocity. |

In the COVID dataset (Fig. 6b), early 2020 exhibits the most prominent clusters of confirmed (green) following relations, particularly from `GeneralChat` toward health-related subtopics such as `MaskUsage`, `LongCovid`, and `AlternativeTreatments`. This temporal clustering corresponds with the outbreak period, when user attention was highly focused and cross-topic engagement peaked.

In contrast, the climate dataset (Fig. 6c) covers a much longer temporal window (2009–2025) and displays several distinct phases of interaction. Between 2010 and 2014, dense clusters of green and grey events occur among `ClimateChange`, `RenewableEnergy`, and `PopulationEconomy`, indicating an early period of sustained cross-topic discussion.

Across domains, a common pattern emerges: large-scale social events or global concerns (e.g., the pandemic, environmental crises, or rapid technological advances) trigger bursts of concentrated engagement that gradually diffuse into smaller, less persistent interaction clusters. The technology and COVID networks show sharper peaks of engagement followed by rapid decline, reflecting event-driven participation. The climate network, by contrast, exhibits longer-term continuity, with engagement sustained across more than a decade but distributed among a broader set of topics.

Together, these results demonstrate that while the inferred following mechanism identifies temporally consistent relationships within each domain, the persistence, density, and thematic breadth of such relationships differ markedly across social contexts. The method thus provides a comparative lens on how online communities organize around short-term versus enduring societal issues.

## 6 Analysis and benchmarking

This section evaluates the inferred social interaction networks quantitatively, benchmarking their structural properties against established graph-theoretic baselines. The aim is to assess the reliability, robustness, and interpretive validity of the inferred relationships—particularly whether observed patterns such as strong hubs, cohesive clusters, and triadic closures reflect genuine structural organization rather than artifacts of data sparsity or noise. Table 4 summarizes the key structural characteristics of inferred interaction networks across topics, highlighting differences in density, reciprocity, and clustering patterns.

### 6.1 Network-level structural metrics

We first analyze global structural properties of the inferred interaction networks across topics. Table 5 summarizes core network-level metrics, including density, clustering coefficient, reciprocity, and average path length. Climate discussions produce the densest and most clustered network, indicating sustained and overlapping conversational interactions among users. The high clustering coefficient suggests frequent triadic closures, consistent with conversational cascades and repeated engagement. COVID-related discussions exhibit a sparse and weakly clustered structure, reflecting episodic or broadcast-style interactions. Technology discussions show low density but moderate clustering, indicating selective engagement around central users or technical experts.

Across all topics, the average path length remains low, suggesting efficient information reach despite differing interaction patterns.

Table 5: Network-level structural metrics across topics

| Metric | Climate | COVID | Technology |
| --- | --- | --- | --- |
| Nodes | 14 | 7 | 33 |
| Edges | 35 | 7 | 40 |
| Density | 0.192 | 0.167 | 0.038 |
| Clustering | 0.765 | 0.295 | 0.349 |
| Reciprocity | 0.286 | 0.000 | 0.000 |
| Avg. Path Length | 1.67 | 1.67 | 1.92 |

## 6.2 Degree distributions and assortativity

Degree statistics reveal substantial heterogeneity across topics. Climate interactions exhibit relatively balanced in- and out-degree distributions, suggesting mutual conversational engagement. In contrast, technology discussions are dominated by a small number of high out-degree users interacting with many passive participants, producing hub-centric structures. COVID discussions show minimal degree variance, consistent with limited interaction depth.

The absence of strong assortative mixing across all topics indicates that users do not preferentially interact with others of similar activity levels. Instead, interaction frequency appears driven by topical salience and participation timing rather than structural homophily in degree space.

## 6.3 Community structure and modularity

We detect communities using modularity maximization and report results in Table 6. Climate discussions form few, weakly separated communities, suggesting a shared conversational space with cross-community engagement. Technology discussions exhibit the highest modularity and the largest number of communities, reflecting topical specialization and segmentation. COVID discussions show limited community formation, primarily due to the small network size rather than strong separation.

These findings suggest that community emergence is topic-dependent, with technical domains encouraging specialization while socio-political topics foster broader interaction.

Table 6: Community structure metrics

| Metric | Climate | COVID | Technology |
| --- | --- | --- | --- |
| Modularity | 0.083 | 0.122 | 0.258 |
| Number of Communities | 3 | 2 | 6 |
| Largest Community Size | 7 | 5 | 19 |

## 6.4 Filter bubbles and echo chamber tendencies

To quantify exposure isolation, we compute a filter bubble metric based on within-community interaction concentration. Across all topics, this metric remains at or near zero, indicating no strong evidence of echo chambers within inferred interaction networks.

This suggests that although communities exist particularly in technology discussions, they are not isolated. Instead, users maintain interaction pathways across community boundaries, likely facilitated by shared discussion threads and temporal overlap.

### 6.5 Robustness and sensitivity analysis

We assess robustness by varying interaction thresholds and temporal persistence requirements used to infer latent followings. Network-level patterns, including clustering trends and degree heterogeneity, remain stable across reasonable parameter ranges. While absolute edge counts fluctuate, relative topic-level differences persist, indicating that observed structural distinctions are not artifacts of threshold selection.

This stability supports the validity of interaction persistence as a signal for latent social ties.

### 6.6 Temporal benchmarking and structural evolution

Finally, we examine how inferred network structure evolves over time. Despite substantial differences in post volume across topics, inferred networks stabilize quickly once sufficient interaction history accumulates. Climate discussions show gradual densification and increasing clustering, while technology discussions stabilize around hub-dominated configurations. COVID discussions show limited temporal evolution due to sparse interaction repetition.

These findings highlight that temporal persistence of interaction—not raw activity volume—is critical for latent network formation.

### 6.7 Cross-domain comparative insights

Comparing inferred interaction networks across domains reveals systematic differences in how conversational attention, engagement persistence, and structural cohesion emerge under varying topical contexts.

Climate discussions consistently exhibit the most cohesive interaction structure, characterized by high clustering, moderate reciprocity, and balanced degree distributions. This suggests sustained multi-turn exchanges and conversational cascades, aligning with issues that invite prolonged debate, norm formation, and collective sense-making. The presence of reciprocal ties further indicates a transition from parasocial attention to stable social engagement.

COVID-related discussions, despite high post volume, generate sparse and weakly connected networks with minimal reciprocity and limited community structure. This pattern reflects episodic engagement driven by external events and information dissemination rather than ongoing dialogue. Interactions are predominantly one-directional, suggesting that users react to content rather than form persistent conversational ties.

Technology discussions occupy an intermediate position, combining low network density with relatively high modularity. A small number of high out-degree users dominate interactions, producing hub-centric structures consistent with expert-driven or influencer-mediated communication. Community segmentation is more pronounced, indicating topical specialization and selective engagement rather than broad conversational mixing.

Across all domains, the absence of strong filter bubble effects suggests that topic-driven participation does not necessarily lead to structural isolation. Instead, interaction persistence and degree heterogeneity jointly shape network evolution in a domain-specific manner.

## 7 Discussion

### 7.1 Topic-Specific Interaction Dynamics

Our results demonstrate that interaction-based latent following networks exhibit strong topic-dependent structural differences, despite being inferred from the same underlying behavioral rule. Climate discussions form dense and highly clustered interaction networks, indicating sustained conversational engagement and repeated attention among users. In contrast, COVID-19 discussions appear sparse and largely non-reciprocal, suggesting broadcast-style or episodic information sharing rather than dialogue. Technology discussions occupy an intermediate regime, characterized by hub-dominated and modular structures that reflect expert-driven or problem-solving interactions. These findings indicate that topics semantics strongly shape how attention and interaction persist over time, even when explicit follow mechanisms are absent.

## 7.2 Information Diffusion and Conversational Cascades

The low average path lengths observed across all topics indicate that information can spread rapidly through inferred interaction networks. However, the mechanism of diffusion differs by topic. In climate discussions, high clustering and moderate reciprocity suggest cascade-like diffusion driven by conversational reinforcement and repeasted exchanges. In technology discussions, diffusion is more centralized with a small number of highly active users initiating interactions with many passive participants. COVID-19 discussions exhibit limited cascade depth, consistent with one-off replies rather than multi-turn conversational chains. This highlights that information flow is not only a function of network connectivity, but also of interaction persistence and reciprocity.

## 7.3 Homophily and community structure

The modularity and community structure further reflect homophilic tendencies that vary by topic. Climate discussions show less modularity and few communities, suggesting that users interact across viewpoints within a shared conversational space. Technology discussions, by contrast, display higher modularity and a larger number of communities, consistent with topical segmentation and specialization. COVID-19 discussions show limited fragmentation, but this is largely driven by the small size of the inferred network rather than strong community structure. These patterns suggest that homophily manifests differently depending on whether a topic encourages debate, information seeking, or broadcasting behavior.

## 7.4 Reciprocity and engagement life cycles

Reciprocity emerges as a key differentiator across topics. Climate discussions exhibit non-zero reciprocity, indicating back-and-forth exchanges and engagement cycles that persist over time. In contrast, the absence of reciprocity in COVID-19 and technology discussions suggests predominantly parasocial or one-directional attention. This aligns with prior work showing that repeated interactions often precede the formation of stable social ties, while one-off interactions rarely evolve into reciprocal relationships.

## 7.5 Filter Bubbles and exposure diversity

Across all topics, the filter bubbles metric remains close to zero, indicating no strong evidence of exposure isolation within inferred communities. This suggests that, at least within the scope of interaction-based following, users are not confined to tightly segregated attention clusters. Instead, observed community structures appear to emerge from topical or behavioral specialization rather than ideological isolation. This findings highlights that filter bubbles may be less pronounced in conversational settings where interaction requires direct engagement, as opposed to passive content consumption.

## 7.6 Implications for Modeling Social Attention

Taken together, these results demonstrate that simple, interpretable rule based on temporal interaction consistency can recover meaningful latent social structures across diverse topics. The inferred networks capture conversational depth, engagement asymmetry, and community organization without relying on explicit social links. This makes the approach particularly suitable for platforms such as Reddit, where following relationships are implicit and topic-dependent.

# 8 Validation and limitations

We validate our inference framework through multiple complementary strategies. First, robustness analyses show that network-level trends remain stable under variations in temporal windowing and interaction thresholds. Second, manual evaluation by human annotators confirms that repeated directed interactions often correspond to meaningful conversational attention and engagement continuity.

Despite these strengths, several limitations remain. The inference of following relationships relies solely on observable reply behavior and does not account for unobserved attention mechanisms such as passive reading or external platform interactions. Temporal aggregation may also obscure short-lived but meaningful interactions, particularly in fast-moving domains such as COVID-related discussions. Furthermore, the relatively small inferred networks—especially in narrower domains—limit the detection of higher-order structural phenomena such as rich-club effects or long-range diffusion pathways.

Finally, while the framework is designed to be platform-agnostic, domain-specific norms and moderation practices may influence interaction visibility and persistence. As a result, inferred following ties should be interpreted as proxies for sustained attention rather than direct equivalents of explicit social connections.

Overall, these findings support the validity of temporal interaction persistence as an interpretable signal for latent social ties, while highlighting the need for cautious interpretation and future extensions incorporating content semantics and richer temporal dynamics.

# 9 Applications and Future Work

The inferred interaction networks and benchmarking framework developed in this study provide a versatile foundation for a range of applied and research-oriented use cases. Beyond describing the structural properties of social discussions, these networks serve as testbeds for evaluating social media dynamics, intervention strategies, and computational models of user behavior.

## 9.1 Applications

The inferred networks enable the simulation and assessment of information interventions, such as moderation policies, content recommendation changes, or targeted awareness campaigns. By modeling how information or influence propagates through empirically grounded structures, researchers can test how modifying connectivity or interaction probabilities might reduce polarization, enhance information diversity, or mitigate misinformation spread. For example, nodes corresponding to bridging topics (*e.g.*, `GeneralChat`, `ScienceTrust`) could be used as insertion points for fact-checking or cross-community content to evaluate their diffusion effects under controlled conditions.

The network structure provides a realistic framework on which to create and study simulated social systems. These environments can be used to study emergent behaviors such as opinion clustering, or viral content diffusion. Because the networks exhibit empirically observed features—small-world structure, and modular clustering—they provide a more realistic base for testing theories of collective behavior, coordination, and conflict formation. Agent-based simulations initialized on these structures can help explore how small changes, such as exposure to new information sources, alter long-term community alignment or fragmentation.

The dataset and derived interaction graphs can further function as a benchmark for evaluating the behavioral realism of social agents or generative dialogue models. By comparing simulated agent interactions against empirical metrics (e.g., reciprocity, clustering, modularity, and degree distributions), researchers can assess whether artificial agents replicate human-like participation patterns or exhibit implausible dynamics. Such benchmarking could guide the development of more sociologically grounded AI systems, ensuring that synthetic communities and conversational agents align with observed social structures.

## 9.2 Future Work

While the current framework provides strong empirical grounding and analytical capability, several directions offer opportunities for enhancement and expansion. Future work will extend the methodology beyond Reddit to platforms such as Twitter/X, Mastodon, and specialized forums. Cross-platform comparison would enable the study of how interaction mechanisms and platform affordances shape network structure, reciprocity, and polarization, offering a more comprehensive view of digital social ecosystems.

Incorporating machine learning models that capture temporal dependencies—for instance, recurrent or dynamic graph neural networks—would allow automated tracking of evolving communities, emerging influencers, and polarization trajectories over time. Modeling cross-community interactions could also reveal how information or users migrate between topics, providing a richer understanding of discourse diffusion and the breakdown (or reinforcement) of filter bubbles.

An important long-term objective is to develop an open simulation environment based on the inferred networks. Such an environment would integrate realistic agent behaviors, communication dynamics, and empirical network structures, serving as a shared research platform for testing interventions, evaluating AI moderation tools, and studying large-scale social phenomena. By releasing these datasets and simulation modules under open standards, future research can systematically benchmark models of collective intelligence, social influence, and misinformation resilience.

Ultimately, integrating these approaches could bridge computational social science and AI research, enabling interpretable, data-driven understanding of online social dynamics. Through continual refinement—adding temporal, semantic, and cross-platform perspectives—the framework can evolve into a unified testbed for studying digital society at scale.

## 10 Conclusions

This work presented the first in combining Reddit-based, agent-aligned dataset designed explicitly for grounding and benchmarking social simulations. By systematically extracting posts, comments, and inferred user-to-user interactions, we constructed a directed, weighted social graph that captures realistic structural and behavioral dynamics across multiple topical domains. Through topic modeling and behavioral clustering, users were categorized into interpretable agent types—allowing the empirical characterization of how different communities interact, follow, and respond to one another.

Our analysis revealed clear empirical regularities: degree distributions followed a heavy-tailed pattern typical of real social networks, while clustering and modularity analyses indicated the presence of well-defined subcommunities. These findings demonstrate that social interactions on Reddit are not randomly distributed but structured around tightly knit, topic-specific clusters that limit cross-domain exposure. Despite these structural constraints, bridging nodes such as `GeneralChat`, `ScienceTrust`, and `ClimateChange` were found to play a disproportionately large role in connecting otherwise fragmented communities—suggesting potential leverage points for intervention or content diversification.

From a modeling perspective, the dataset provides a robust empirical foundation for both traditional and LLM-driven agent-based models. It enables simulation designers to calibrate behavioral rules, validate diffusion outcomes, and benchmark emergent dynamics against real-world interaction baselines. The results demonstrate that data-grounded simulation frameworks can reproduce observed engagement and connectivity patterns, marking a step toward socially realistic, empirically validated agent societies.

In summary, this study contributes (1) a large-scale, empirically derived dataset linking user content and interaction structure on Reddit, (2) a methodological pipeline for constructing and analyzing social networks suitable for simulation calibration, and (3) empirical insights into online community organization, engagement asymmetries, and filter bubble formation. Together, these elements establish a clear path toward data-grounded, reproducible, and socially valuable agent-based models—bridging the gap between synthetic simulation and real human behavior in online ecosystems.

# 11 Acknowledgment

This work is supported by TWON (project number 101095095), a research project funded by the European Union under the Horizon Europe framework (HORIZON-CL2-2022-DEMOCRACY-01, Topic 07). The authors acknowledge the use of artificial intelligence tools in the preparation of this manuscript. Specifically, ChatGPT (OpenAI) was used to assist with drafting and refining portions of the text and with providing explanations related to software implementation. All AI-generated content was reviewed, validated, and approved by the authors, who take full responsibility for the accuracy and integrity of the manuscript.

# 12 Conflict of Interest

The authors declare that they have no known competing financial interests or personal relationships that could have appeared to influence the work reported in this paper.

# 13 Appendix

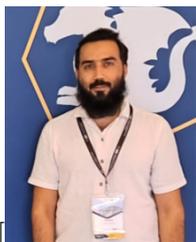Abdul Sittar is a Postdoctoral Researcher specializing in machine learning and natural language processing (NLP), with over a decade of expertise in the field. His work focuses on developing advanced models using frameworks like PyTorch, TensorFlow, and Hugging Face. He actively collaborate with academia, contributing to both theoretical and applied research, with a strong record of publications in peer-reviewed international journals. He was a MSCA fellow working on CLEOPATRA ITN. He acquired MS degree from COMSATS Institute of Information Technology, Lahore, Pakistan.

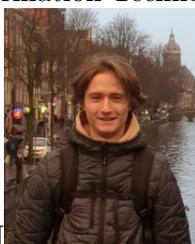Miha is a Master's student in Mathematics and Computer Science at the Faculty of Mathematics and Physics (FMF), University of Ljubljana. He works as a student researcher at the Jožef Stefan Institute (JSI), where he assists in research on project TWON. He has completed his Bachelor's degree in Mathematics and Computer Science and is interested in the application of mathematical and computational approaches to real-world research problems.

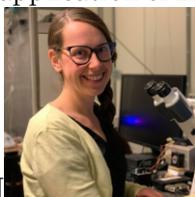Alenka Guček is a professional research associate at Department of Artificial Intelligence at Jozef Stefan Institute. She has a PhD from Biosciences at University of Ljubljana, Slovenia and she worked as a postdoc in diabetes and molecular cell biology at Uppsala University in Sweden, where she still teaches data visualization. Her current work at AI department of JSI includes codesign of exploratory data visualization tools, observatories and AI services that meet the needs of users.

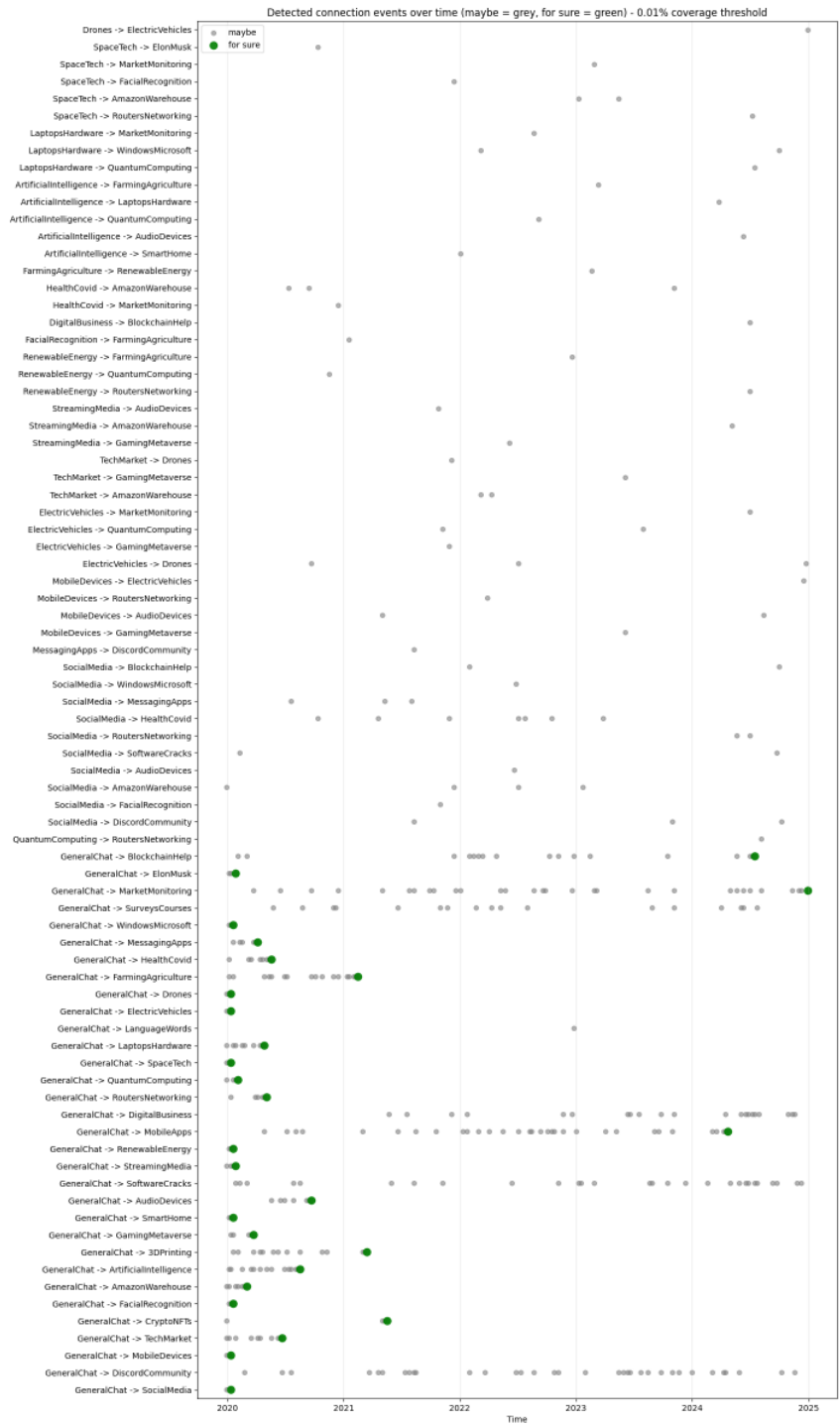

Figure 7: Technology domain: detected following events over time at 0.01% threshold.

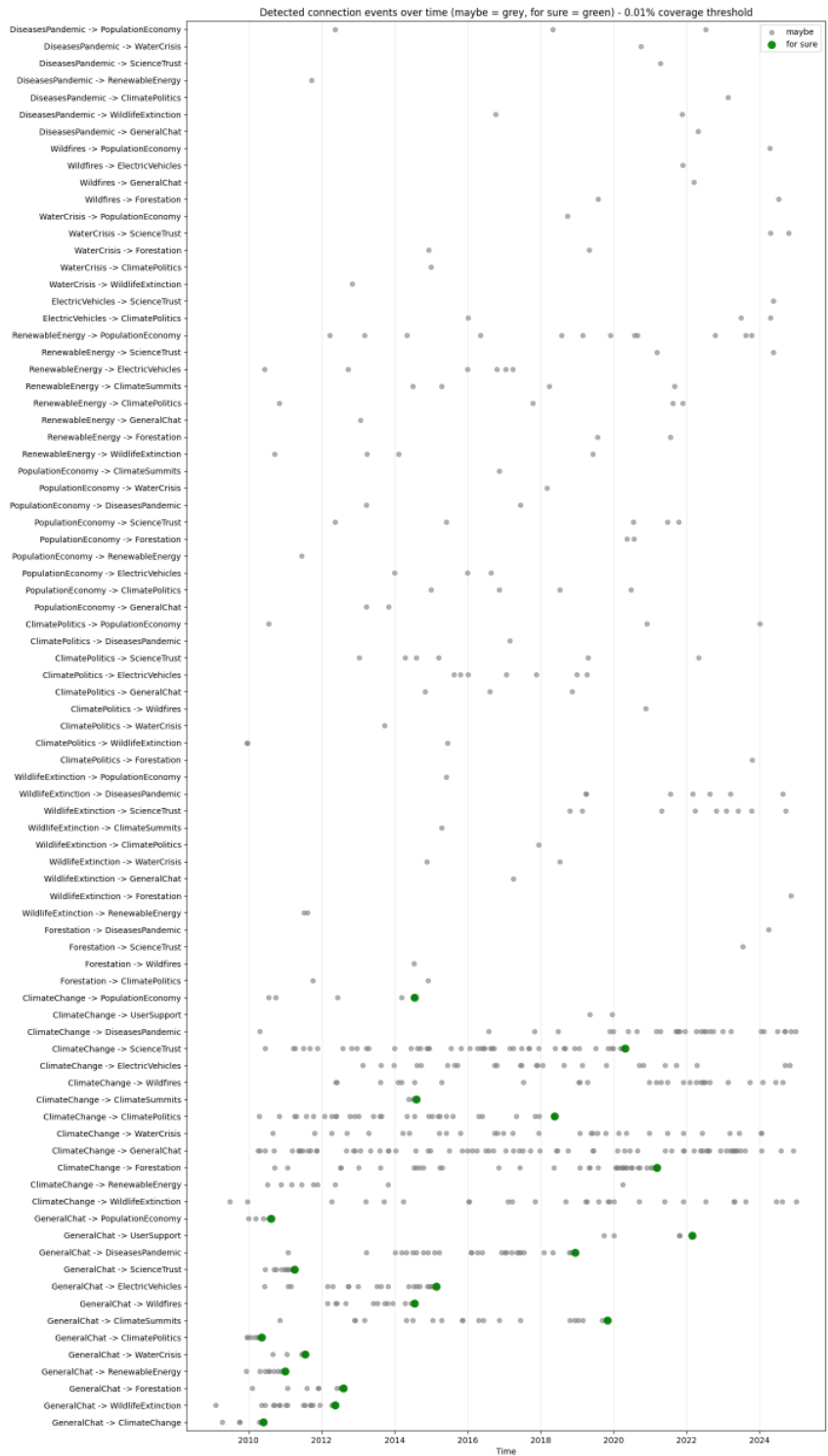

Figure 8: Climate domain: detected following events over time at 0.01% threshold.

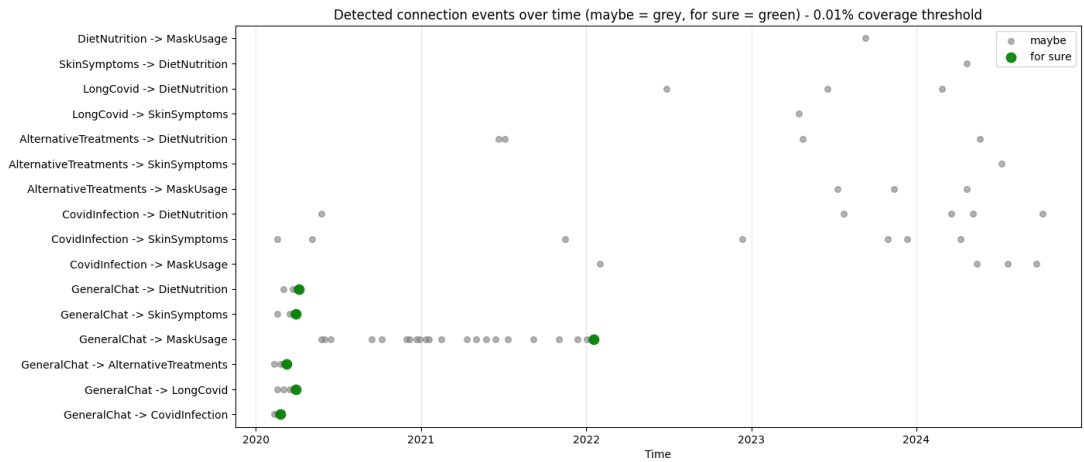

Figure 9: COVID domain: detected following events over time at 0.01% threshold.

Figure 10: Qualitative evaluation (example 1): Cascade analysis, engagement life cycle, homophily, sentiment analysis, conversational dynamics, topic drifting

Figure 11: Qualitative evaluation (example 2): Cascade analysis, engagement life cycle, homophily, sentiment analysis, conversational dynamics, topic drifting

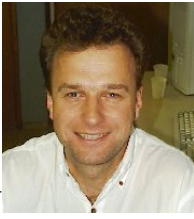Marko Grobelnik is an expert researcher in the field of Artificial Intelligence (AI). Focused areas of expertise are Machine Learning, Data/Text/Web Mining, Network Analysis, Semantic Technologies, Deep Text Understanding, and Data Visualization. Marko co-leads the Department for Artificial Intelligence at Jozef Stefan Institute, co-founded UNESCO International Research Center on AI (IRCAI), and is the CEO of Quintelligence.com specialized in solving complex AI tasks for the commercial world. He collaborates with major European academic institutions and major industries such as Bloomberg, British Telecom, European Commission, Microsoft Research, New York Times. Marko is co-author of several books, co-founder of several start-ups and is/was involved into over 50 EU funded research projects in various fields of Artificial Intelligence.